\documentclass[journal]{IEEEtran}
\IEEEoverridecommandlockouts

\usepackage{amsmath,amssymb,amsfonts}
\usepackage{algorithmic}
\usepackage{graphicx}
\usepackage{textcomp}
\usepackage{xcolor}
\usepackage{cite}
\usepackage{balance}
\usepackage{microtype}
\usepackage{url}
\usepackage[compatibility=false]{caption}
\usepackage{subcaption}

\def\BibTeX{{\rm B\kern-.05em{\sc i\kern-.025em b}\kern-.08em
    T\kern-.1667em\lower.7ex\hbox{E}\kern-.125emX}}

\usepackage{pgfplots}
\pgfplotsset{compat=1.18}
\usepgfplotslibrary{groupplots}
\usetikzlibrary{patterns,calc}
\definecolor{baseGrey}{HTML}{d9aa00}
\definecolor{caeBlue}{HTML}{8DA6CC}
\pgfplotsset{
  caestyle/.style={
    ybar,
    ymajorgrids=true,
    grid style={line width=0.3pt, draw=gray!30},
    axis lines*=left,
    tick label style={font=\footnotesize},
    label style={font=\footnotesize},
    ylabel near ticks,
    xtick=data,
    enlarge x limits=0.35,
  }
}

\usetikzlibrary{arrows.meta,positioning,decorations.pathreplacing}
\tikzset{
  pf/.style={line width=0.6pt, font=\scriptsize},
  pftitle/.style={font=\footnotesize, anchor=base},
  pflabel/.style={font=\scriptsize},
  pfarrow/.style={-{Straight Barb[length=3pt,width=4pt]}, line width=0.5pt},
  pfdash/.style={dash pattern=on 2.5pt off 2pt, line width=0.6pt},
  pfdasharrow/.style={pfdash, -{Straight Barb[length=3pt,width=4pt]}},
  pfframe/.style={draw, line width=0.5pt, fill=white},
  pfsel/.style={draw, line width=0.6pt, fill=gray!55},
  pfnode/.style={circle, draw, line width=0.5pt, minimum size=5pt, inner sep=0pt, fill=white},
  pfgrey/.style={circle, draw=gray!70!black, line width=0.4pt, minimum size=6pt, inner sep=0pt, fill=gray!55},
  pfblack/.style={circle, fill=black, minimum size=6pt, inner sep=0pt},
  pffrozen/.style={draw, line width=0.6pt, pattern=north east lines, pattern color=gray!60},
  pfthin/.style={line width=0.3pt, draw=gray!70},
}

\newcommand{\pfpair}[5]{%
  \pgfmathsetmacro{\wA}{#4/6}
  \pgfmathsetmacro{\wB}{#4/8}
  \foreach \i in {1,...,6}{
    \ifnum\i=3 \def\sty{pfsel}\else\def\sty{pfframe}\fi
    \draw[\sty] ({#1+(\i-1)*\wA},#2) rectangle ++(\wA,#5);
  }
  \foreach \j in {1,...,8}{
    \ifnum\j=4 \def\sty{pfsel}\else\def\sty{pfframe}\fi
    \draw[\sty] ({#1+(\j-1)*\wB},#3) rectangle ++(\wB,#5);
  }
  \foreach \i/\j in {1/1,2/2,2/3,3/4,4/5,5/6,5/7,6/8}{
    \ifnum\j=4 \def\lw{0.9pt}\def\lc{black}\else\def\lw{0.35pt}\def\lc{black!70}\fi
    \draw[line width=\lw, draw=\lc] ({#1+(\i-0.5)*\wA},#2) -- ({#1+(\j-0.5)*\wB},{#3+#5});
  }
  \coordinate (xa-top) at ({#1+2.5*\wA},{#2+#5});
  \coordinate (xa-c)   at ({#1+2.5*\wA},{#2+#5/2});
  \coordinate (xb-bot) at ({#1+3.5*\wB},#3);
  \coordinate (xb-c)   at ({#1+3.5*\wB},{#3+#5/2});
}

\usepackage{booktabs}
\usepackage{caption}
\usepackage{tabularx}
\usepackage{multirow}
\newcommand{\mytable}{
    \centering
    \small
    \renewcommand{\arraystretch}{1.2}
    }

\newcolumntype{C}{>{\centering\arraybackslash}X}
\newcolumntype{L}{>{\raggedright\arraybackslash}X}

\usepackage{amsfonts}
\usepackage{bm}

\usepackage{acro}
\DeclareAcronym{ssl}{short=SSL, long=self-supervised learning}
\DeclareAcronym{cae}{short=cAE, long=correspondence autoencoder}
\DeclareAcronym{utd}{short=UTD, long=unsupervised term discovery}
\DeclareAcronym{ap}{short=AP, long=average precision}
\DeclareAcronym{mfcc}{short=MFCC, long = mel-frequency cepstral coefficient}
\DeclareAcronym{bnf}{short=BNF, long = bottleneck features}
\DeclareAcronym{dtw}{short=DTW, long = dynamic time warping}
\DeclareAcronym{ae}{short=AE, long= autoencoder}
\DeclareAcronym{sw}{short= SW, long = same word}
\DeclareAcronym{prb}{short = PRB, long = precision-recall break-even point}
\DeclareAcronym{swsp}{short = SWSP, long= same word same speaker}
\DeclareAcronym{swdp}{short = SWDP, long= same word different speaker}
\DeclareAcronym{dwsp}{short = DWSP, long= different word same speaker}
\DeclareAcronym{dwdp}{short = DWDP, long= different word different speaker}
\DeclareAcronym{cmvn}{short=CMVN, long= cepstral mean and variance normalisation}

\begin{document}

\title{Lightweight Adaptation of Self-Supervised Features using Discovered Spoken-Term Pairs}

\author{Camryn Abrahamson, Nicol Visser, and Herman Kamper
\thanks{C. Abrahamson, N. Visser, and H. Kamper are with the Department of Electrical and Electronic Engineering, Stellenbosch University, South Africa (email: 25900048@sun.ac.za, 16986431@sun.ac.za, kamperh@sun.ac.za).}}

\markboth{Submitted to IEEE Signal Processing Letters}{Abrahamson \MakeLowercase{\textit{et al.}}: Lightweight Adaptation of Self-Supervised Features using Discovered Spoken-Term Pairs}

\maketitle

\begin{abstract}
Features from self-supervised speech models such as WavLM degrade when applied to speech from unseen domains or languages, and common adaptation methods require transcriptions or backpropagation through large models.
We instead keep the backbone model frozen and train a lightweight network on top of it, using pairs of spoken terms discovered without supervision.
We compare two lightweight models: the correspondence autoencoder~(cAE), which tries to construct a frame from one term given a frame from another as input, and a noise contrastive estimation (NCE) model, which maximises the frame similarity of paired terms relative to negatives.
On English, Xitsonga, and French data, both models improve phone ABX error and word same-different average precision~(AP) over the WavLM baseline, with AP gains coming mainly from different-speaker comparisons.
The cAE gives better ABX scores on English and Xitsonga, while the NCE model gives better AP on Xitsonga and French.
Compared to SpidR-Adapt, a recent method that fine-tunes the model itself, our approach performs better on English and competitively on Xitsonga, but has worse ABX
on French.
\end{abstract}

\begin{IEEEkeywords}
correspondence autoencoder, domain adaptation, self-supervised speech models, unsupervised term discovery
\end{IEEEkeywords}

\IEEEpeerreviewmaketitle

\section{Introduction}

\IEEEPARstart{S}{peech} models employing self-supervised learning (SSL) such as WavLM~\cite{chen+etal_jstsp22} and HuBERT~\cite{hsu+etal_taslp21} have become integral in speech processing.
These models are pretrained on large amounts of predominantly English speech, and their performance drops considerably on speech that differs from the pretraining data in language
or speaking-style~\cite{hsu+etal_interspeech21, smit2025fewshot}.

Several methods for adapting a model to a target domain have been proposed.
Continued pretraining or fine-tuning can mitigate the drop in performance~\cite{sanroman+etal_arxiv26}, but requires substantial target-domain audio and backpropagation through the full model.
Parameter-efficient methods reduce the number of trained parameters~\cite{thomas+etal_icassp22, hu+etal_iclr22}, but still backpropagate through the model.
Test-time adaptation~\cite{lin+etal_interspeech22} instead optimises for a specific downstream task rather than improving the representation itself.
At the lightest end, feature-space methods~\cite{viikki+laurila_speechcom98, sun+etal_aaai16, vanniekerk+etal_interspeech21} require no training, but are limited to simple statistical transformations.
Closest to our setting,
\cite{meghanani+hain_icassp24} applies correspondence training using pairs of original and augmented speech, and SpidR-Adapt~\cite{luthra+etal_acl26} meta-learns an initialisation for adaptation from small amounts of audio, but both fine-tune the full model itself.

In this work, we keep the SSL model frozen and train a small network on its features, using pairs of terms discovered without transcriptions through a recent unsupervised term discovery system called DUSTED~\cite{vanniekerk+etal_interspeech24}.
We build on the frame-level correspondence autoencoder
(cAE)~\cite{kamper+etal_icassp15,renshaw+etal_interspeech15}, which tries to reconstruct one frame in a discovered pair given another as input.
Although correspondence training has been applied to SSL representations for segment-level acoustic embeddings~\cite{meghanani+hain_eacl24,whealing+etal_slt26}, the cAE has not been evaluated as a lightweight adaptation method of frame-level SSL features.
As an alternative, we consider a model trained with noise contrastive estimation (NCE)~\cite{gutmann+hyvarinen_aistats10,oord+etal_arxiv18}, where paired frames are explicitly pulled together while mismatched frames are pushed apart.
Previously, on MFCCs, the cAE outperformed an older contrastive model~\cite{last+etal_spl20}, but neither has been tested as an unsupervised adaptation method for modern SSL features.

On English, Xitsonga, and French data, we find that unsupervised adaptation using discovered terms improve both phone-
and word-level discrimination performance over the SSL baseline~\cite{chen+etal_jstsp22} in all three languages.
The cAE achieves better phone-level discrimination on English and Xitsonga while NCE performs better at the word level on Xitsonga and French.
Compared to SpidR-Adapt~\cite{luthra+etal_acl26}, our approach performs better on English and competitively on Xitsonga, but has higher phone-level error on French.

Our contributions are:
(1) a lightweight adaptation approach that
requires no transcriptions by using a new term discovery approach;
(2) the first evaluation of the cAE for adapting frame-level SSL features, compared to a contrastive alternative; and
(3) results on three languages showing gains over WavLM and competitive performance with full-model fine-tuning.
Our code is available at \url{https://github.com/camrynx/correspondence/}.

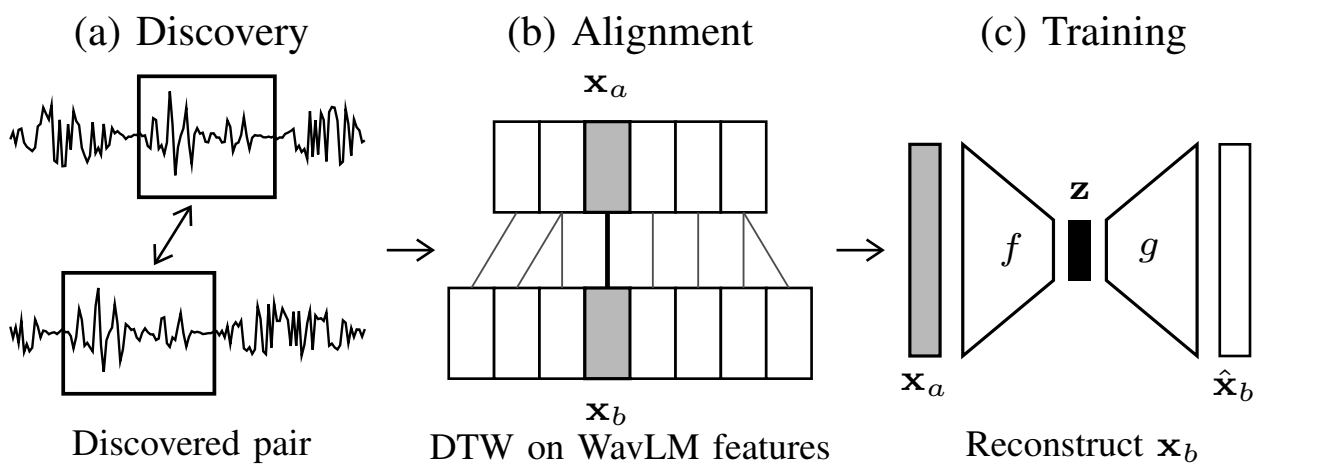
\begin{figure}[!b]
  \centering
  \begin{tikzpicture}[pf]
\node[pftitle, anchor=south, inner ysep=1pt] at (1.2, 2.5) {(a) Discovery};
  \node[pftitle, anchor=south, inner ysep=1pt] at (4.1, 2.5) {(b) Alignment};
  \node[pftitle, anchor=south, inner ysep=1pt] at (7.1, 2.5) {(c) Training};

  \draw[line width=0.45pt] (0.000,1.983) -- (0.018,2.030) -- (0.036,1.954) -- (0.055,2.046) -- (0.073,2.054) -- (0.091,1.962) -- (0.109,1.934) -- (0.127,1.920) -- (0.145,1.905) -- (0.164,1.928) -- (0.182,2.069) -- (0.200,2.085) -- (0.218,2.193) -- (0.236,2.153) -- (0.255,2.114) -- (0.273,2.210) -- (0.291,1.842) -- (0.309,1.828) -- (0.327,2.112) -- (0.345,1.805) -- (0.364,1.807) -- (0.382,2.083) -- (0.400,1.915) -- (0.418,2.172) -- (0.436,2.104) -- (0.455,2.077) -- (0.473,1.927) -- (0.491,1.877) -- (0.509,1.833) -- (0.527,1.875) -- (0.545,2.097) -- (0.564,1.891) -- (0.582,1.926) -- (0.600,1.881) -- (0.618,2.085) -- (0.636,1.924) -- (0.655,1.946) -- (0.673,2.058) -- (0.691,2.023) -- (0.709,2.037) -- (0.727,2.020) -- (0.745,2.009) -- (0.764,1.985) -- (0.782,2.001) -- (0.800,2.000) -- (0.818,2.013) -- (0.836,2.011) -- (0.855,2.026) -- (0.873,2.004) -- (0.891,2.009) -- (0.909,1.939) -- (0.927,1.923) -- (0.945,2.123) -- (0.964,2.111) -- (0.982,1.920) -- (1.000,1.944) -- (1.018,1.885) -- (1.036,1.962) -- (1.055,2.300) -- (1.073,2.082) -- (1.091,1.745) -- (1.109,1.931) -- (1.127,2.006) -- (1.145,2.056) -- (1.164,2.192) -- (1.182,1.953) -- (1.200,1.819) -- (1.218,2.017) -- (1.236,2.057) -- (1.255,2.020) -- (1.273,2.027) -- (1.291,1.967) -- (1.309,1.963) -- (1.327,2.022) -- (1.345,2.023) -- (1.364,1.995) -- (1.382,1.995) -- (1.400,1.962) -- (1.418,1.966) -- (1.436,2.109) -- (1.455,2.066) -- (1.473,1.901) -- (1.491,1.953) -- (1.509,1.962) -- (1.527,2.012) -- (1.545,2.130) -- (1.564,2.002) -- (1.582,1.918) -- (1.600,1.994) -- (1.618,2.011) -- (1.636,2.007) -- (1.655,2.009) -- (1.673,1.997) -- (1.691,1.996) -- (1.709,2.001) -- (1.727,1.995) -- (1.745,2.004) -- (1.764,1.984) -- (1.782,1.996) -- (1.800,1.994) -- (1.818,2.019) -- (1.836,2.027) -- (1.855,2.038) -- (1.873,2.052) -- (1.891,1.914) -- (1.909,1.941) -- (1.927,1.899) -- (1.945,1.919) -- (1.964,1.852) -- (1.982,2.169) -- (2.000,1.867) -- (2.018,2.110) -- (2.036,1.883) -- (2.055,2.157) -- (2.073,2.112) -- (2.091,1.837) -- (2.109,2.164) -- (2.127,2.197) -- (2.145,2.193) -- (2.164,2.107) -- (2.182,1.818) -- (2.200,1.835) -- (2.218,2.184) -- (2.236,2.085) -- (2.255,1.875) -- (2.273,2.116) -- (2.291,2.063) -- (2.309,2.042) -- (2.327,1.964) -- (2.345,1.976);
  \draw[line width=0.45pt] (0.000,0.744) -- (0.018,0.670) -- (0.036,0.740) -- (0.055,0.635) -- (0.073,0.784) -- (0.091,0.576) -- (0.109,0.629) -- (0.127,0.581) -- (0.145,0.605) -- (0.164,0.791) -- (0.182,0.745) -- (0.200,0.672) -- (0.218,0.718) -- (0.236,0.713) -- (0.255,0.694) -- (0.273,0.703) -- (0.291,0.707) -- (0.309,0.695) -- (0.327,0.711) -- (0.345,0.728) -- (0.364,0.707) -- (0.382,0.709) -- (0.400,0.682) -- (0.418,0.600) -- (0.436,0.677) -- (0.455,0.860) -- (0.473,0.791) -- (0.491,0.617) -- (0.509,0.637) -- (0.527,0.607) -- (0.545,0.580) -- (0.564,0.878) -- (0.582,1.000) -- (0.600,0.618) -- (0.618,0.445) -- (0.636,0.653) -- (0.655,0.706) -- (0.673,0.738) -- (0.691,0.893) -- (0.709,0.765) -- (0.727,0.515) -- (0.745,0.594) -- (0.764,0.761) -- (0.782,0.745) -- (0.800,0.722) -- (0.818,0.728) -- (0.836,0.673) -- (0.855,0.652) -- (0.873,0.705) -- (0.891,0.733) -- (0.909,0.708) -- (0.927,0.693) -- (0.945,0.693) -- (0.964,0.655) -- (0.982,0.665) -- (1.000,0.798) -- (1.018,0.806) -- (1.036,0.634) -- (1.055,0.610) -- (1.073,0.673) -- (1.091,0.656) -- (1.109,0.748) -- (1.127,0.833) -- (1.145,0.702) -- (1.164,0.612) -- (1.182,0.677) -- (1.200,0.714) -- (1.218,0.705) -- (1.236,0.710) -- (1.255,0.705) -- (1.273,0.695) -- (1.291,0.697) -- (1.309,0.701) -- (1.327,0.701) -- (1.345,0.710) -- (1.364,0.715) -- (1.382,0.687) -- (1.400,0.716) -- (1.418,0.674) -- (1.436,0.748) -- (1.455,0.628) -- (1.473,0.747) -- (1.491,0.773) -- (1.509,0.746) -- (1.527,0.783) -- (1.545,0.600) -- (1.564,0.585) -- (1.582,0.608) -- (1.600,0.564) -- (1.618,0.593) -- (1.636,0.879) -- (1.655,0.796) -- (1.673,0.581) -- (1.691,0.562) -- (1.709,0.893) -- (1.727,0.605) -- (1.745,0.896) -- (1.764,0.782) -- (1.782,0.782) -- (1.800,0.864) -- (1.818,0.825) -- (1.836,0.788) -- (1.855,0.616) -- (1.873,0.843) -- (1.891,0.617) -- (1.909,0.847) -- (1.927,0.553) -- (1.945,0.623) -- (1.964,0.767) -- (1.982,0.848) -- (2.000,0.591) -- (2.018,0.819) -- (2.036,0.794) -- (2.055,0.800) -- (2.073,0.782) -- (2.091,0.838) -- (2.109,0.618) -- (2.127,0.632) -- (2.145,0.607) -- (2.164,0.593) -- (2.182,0.622) -- (2.200,0.786) -- (2.218,0.770) -- (2.236,0.634) -- (2.255,0.735) -- (2.273,0.675) -- (2.291,0.726) -- (2.309,0.725) -- (2.327,0.678) -- (2.345,0.728);
  \draw[line width=0.7pt] (0.85,1.6) rectangle (1.75,2.4);
  \draw[line width=0.7pt] (0.35,0.3) rectangle (1.35,1.1);
  \draw[pfarrow, {Straight Barb[length=3pt,width=4pt]}-{Straight Barb[length=3pt,width=4pt]}] (1.2,1.55) -- (0.95,1.15);
  \node[pflabel] at (1.2,-0.05) {Discovered pair};

  \draw[pfframe] (3.20,1.5) rectangle ++(0.3,0.6);
  \draw[pfframe] (3.50,1.5) rectangle ++(0.3,0.6);
  \draw[pfsel] (3.80,1.5) rectangle ++(0.3,0.6);
  \draw[pfframe] (4.10,1.5) rectangle ++(0.3,0.6);
  \draw[pfframe] (4.40,1.5) rectangle ++(0.3,0.6);
  \draw[pfframe] (4.70,1.5) rectangle ++(0.3,0.6);
  \draw[pfframe] (2.90,0.4) rectangle ++(0.3,0.6);
  \draw[pfframe] (3.20,0.4) rectangle ++(0.3,0.6);
  \draw[pfframe] (3.50,0.4) rectangle ++(0.3,0.6);
  \draw[pfsel] (3.80,0.4) rectangle ++(0.3,0.6);
  \draw[pfframe] (4.10,0.4) rectangle ++(0.3,0.6);
  \draw[pfframe] (4.40,0.4) rectangle ++(0.3,0.6);
  \draw[pfframe] (4.70,0.4) rectangle ++(0.3,0.6);
  \draw[pfframe] (5.00,0.4) rectangle ++(0.3,0.6);
  \draw[line width=0.35pt, draw=black!70] (3.35,1.5) -- (3.05,1.0);
  \draw[line width=0.35pt, draw=black!70] (3.65,1.5) -- (3.35,1.0);
  \draw[line width=0.35pt, draw=black!70] (3.65,1.5) -- (3.65,1.0);
  \draw[line width=0.9pt, draw=black] (3.95,1.5) -- (3.95,1.0);
  \draw[line width=0.35pt, draw=black!70] (4.25,1.5) -- (4.25,1.0);
  \draw[line width=0.35pt, draw=black!70] (4.55,1.5) -- (4.55,1.0);
  \draw[line width=0.35pt, draw=black!70] (4.85,1.5) -- (4.85,1.0);
  \draw[line width=0.35pt, draw=black!70] (4.85,1.5) -- (5.15,1.0);
  \coordinate (xa-top) at (3.95,2.1);
  \coordinate (xb-bot) at (3.95,0.4);
  \node[pflabel, above=1pt] at (xa-top) {$\mathbf{x}_a$};
  \node[pflabel, below=1pt] at (xb-bot) {$\mathbf{x}_b$};
  \node[pflabel] at (4.1,-0.05) {DTW on WavLM features};
  \draw[pfarrow] (2.49,1.25) -- (2.81,1.25);

  \draw[pfsel] (5.95,0.55) rectangle (6.15,1.95);
  \draw[line width=0.6pt, fill=white] (6.3,0.55) -- (6.3,1.95) -- (6.9,1.45) -- (6.9,1.05) -- cycle;
  \fill (7.0,1.05) rectangle (7.15,1.45);
  \draw[line width=0.6pt, fill=white] (7.25,1.05) -- (7.25,1.45) -- (7.85,1.95) -- (7.85,0.55) -- cycle;
  \draw[pfframe] (8.0,0.55) rectangle (8.2,1.95);
  \node at (6.62,1.25) {$f$};
  \node at (7.53,1.25) {$g$};
  \node[pflabel] at (6.05,0.36) {$\mathbf{x}_a$};
  \node[pflabel] at (7.075,1.64) {$\mathbf{z}$};
  \node[pflabel] at (8.1,0.36) {$\hat{\mathbf{x}}_b$};
  \node[pflabel] at (7.1,-0.05) {Reconstruct $\mathbf{x}_b$};
  \draw[pfarrow] (5.465,1.25) -- (5.785,1.25);
\end{tikzpicture}
  \caption{(a) The DUSTED~\cite{vanniekerk+etal_interspeech24} unsupervised term discovery system finds pairs of spoken terms.
  (b) The terms are aligned with DTW over frozen WavLM features.
  (c) A lightweight network is trained using the discovered frame pairs.}
  \label{fig:method}
\end{figure}

\section{Method}
Given a pretrained SSL model and a small amount of unlabelled speech from a target domain, we aim to obtain representations that are better suited to that domain without updating the SSL model itself.
Our approach consists of three stages, illustrated in Figure~\ref{fig:method}.

First, an unsupervised term discovery (UTD) system~\cite{park+glass_taslp08,vanniekerk+etal_interspeech24} is applied to the target-domain speech to discover pairs of word-like spoken terms.
These terms serve as a supervision for our models, but can be acquired without any transcriptions.
Second, each discovered pair is aligned using dynamic time warping (DTW)~\cite{sakoe+chiba_taslp78} over its SSL features, yielding a set of frame-level pairs.
Third, a lightweight feedforward network is trained on these frames.
We consider two adaptation approaches:
correspondence training, where we
reconstruct one frame in the pair given the other as input, or through contrastive learning.

\subsection{Correspondence autoencoder}

Let $(\mathbf{x}_a, \mathbf{x}_b)$ denote a pair of SSL features obtained through DTW alignment of a discovered word pair (from a UTD system).
Rather than reconstructing its input, as in a standard autoencoder, the correspondence autoencoder (cAE)~\cite{kamper+etal_icassp15}
is trained to reconstruct the corresponding frame from the other instance, in both directions:
\begin{equation}
  \mathcal{L} = \lVert \mathbf{x}_b - g(f(\mathbf{x}_a)) \rVert^2
              + \lVert \mathbf{x}_a - g(f(\mathbf{x}_b)) \rVert^2,
  \label{eq:cae}
\end{equation}
where $f$ and $g$ denote the feedforward encoder and decoder, respectively.
The idea is that the target frame would differ in properties like channel noise and who the speaker is (which we want to remove) but will have the same linguistic content (which is what we want to capture in our features).
The cAE can therefore be viewed as a denoising autoencoder~\cite{vincent+etal_icml08}, where the ``noise'' is not synthetic, but rather natural.

The encoder terminates in a bottleneck layer, $\mathbf{z} = f(\mathbf{x})$, which we take as the adapted representation.

\subsection{Noise contrastive estimation}

An alternative
is to use noise contrastive estimation (NCE) to explicitly maximise a similarity between paired frames while minimising similarity between mismatched frames.
For a frame-pair $(\mathbf{x}_a, \mathbf{x}_b)$, let $\mathbf{z}_a = f(\mathbf{x}_a)$ and $\mathbf{z}_b = f(\mathbf{x}_b)$ denote the output of a feedforward network.
A batch contains $B$ such pairs, giving a set $\mathcal{Z}$ of $2B$ encoded frames.
Using the InfoNCE loss~\cite{oord+etal_arxiv18}, $\mathbf{z}_a$ is pulled towards its paired frame $\mathbf{z}_b$ and pushed away from the other frames in the batch.
Frames originating from the same discovered term pair as $\mathbf{z}_a$, denoted $\mathcal{S}_a \subset \mathcal{Z}$, are excluded as negatives, since these may share linguistic content.
Let $\mathcal{N}_a = \mathcal{Z} \setminus (\mathcal{S}_a \cup \{\mathbf{z}_b\})$ denote the remaining negatives for anchor $\mathbf{z}_a$.
The loss for anchor $\mathbf{z}_a$ is then
\begin{equation}
    \mathcal{L} = -\log
    \frac{e^{\mathrm{sim}(\mathbf{z}_a, \mathbf{z}_b)/\tau}}
       {e^{\mathrm{sim}(\mathbf{z}_a, \mathbf{z}_b)/\tau}
        + \sum_{\mathbf{z} \in \mathcal{N}_a}
          e^{\mathrm{sim}(\mathbf{z}_a, \mathbf{z})/\tau}},
  \label{eq:nce}
\end{equation}
where $\mathrm{sim}(\cdot,\cdot)$ is cosine similarity and $\tau$ is a temperature parameter.
This loss is averaged over all $2B$ frames in the batch as anchors, so that each pair contributes in both directions, as in~\eqref{eq:cae}.
As with the cAE, the encoder output $\mathbf{z}$ is used as the adapted representation.
A similar objective was used to learn segment-level embeddings for zero-resource languages in~\cite{jacobs+etal_slt21}.

\subsection{Finding pairs}
To discover term pairs in speech without transcriptions, we use DUSTED~\cite{vanniekerk+etal_interspeech24}, a state-of-the-art UTD system based on discrete SSL units.\footnote{In development experiments, DUSTED consistently outperformed the original UTD system of~\cite{jansen+vandurme2011} used in~\cite{kamper+etal_icassp15}, which employed classic features.}
DUSTED encodes speech into units by discretising layer-7 features of HuBERT Base~\cite{hsu+etal_taslp21} with $k$-means, and then searches for pairs of similar unit subsequences.
We use the English HuBERT checkpoint for all datasets, making it useable for any language.
But we retrain the $k$-means codebook on the non-English target-domain audio for each case.
In all other respects we retain the default settings.
Pairs in which either term is shorter than 0.5\,s are discarded to reduce training cost.
Each remaining pair is aligned with DTW using cosine distance over the SSL features, yielding a set of frame-pairs that can be used to train correspondence or NCE models.

As a supervised reference, we also train on gold pairs: word tokens of the same type from forced alignments, subject to the same duration filter and requiring at least 5 characters.

\section{Experimental setup}
\subsection{Data}
We perform experiments on three languages.
The datasets are summarised in Table~\ref{tab:data}.
For English, we use the Buckeye corpus of conversational
speech~\cite{pitt+etal_buckeye07}.
Xitsonga is a Bantu language spoken in southern Africa with little publicly available transcribed speech; we use read speech from the NCHLT Xitsonga corpus~\cite{devries+etal_speechcom14}, specifically the data from the ZeroSpeech 2015 challenge~\cite{versteegh+etal_interspeech15}.
For French, we use a subset of the ZeroSpeech 2017 data~\cite{dunbar+etal_asru17}, consisting of read audiobook speech from LibriVox (\url{https://librivox.org}).

The cAE hyperparameters and the NCE temperature are tuned on the English development set and used unchanged for the other two languages.

\begin{table}[!t]
    \newcommand{\twoline}[2]{\begin{tabular}[c]{@{}l@{}}#1\\#2\end{tabular}}
    \mytable
    \footnotesize
    \caption{Duration of active speech, number of speakers, and number of gold and DUSTED-discovered terms (after filtering).}
    \begin{tabularx}{\linewidth}{@{}l@{\ \ \ } l cc@{\ \ \ }C@{\ \ }C@{}}
        \toprule
        Language & Split & Hours & Speakers & Gold terms & Discovered terms \\
        \midrule
        \multirow{3}{*}{\twoline{English}{(Buckeye)}}
            & Train & 5.98 & 12 & 5,053 & 20,366 \\
            & Dev   & 3.55 & \hphantom{0}8  & 2,697 & --- \\
            & Test  & 4.56 & 12 & 4,002 & --- \\
        \midrule
        \multirow{2}{*}{\twoline{Xitsonga}{(NCHLT)}}
            & Train & 1.20 & 12 & 3,135 & 26,840 \\
            & Test  & 1.28 & 12 & 3,249 & --- \\
        \midrule
        \multirow{2}{*}{\twoline{French}{(ZeroSpeech 2017)}}
            & Train & 2.28 & 10 & 4,995 & \hphantom{0}7,740 \\
            & Test  & 1.96 & 17 & 4,064 & --- \\
        \bottomrule
    \end{tabularx}
    \label{tab:data}
\end{table}

\subsection{Evaluation}
We evaluate the adapted features at both the phone and word level.
For the phone level, we use the triphone ABX task~\cite{schatz+etal_interspeech13,versteegh+etal_interspeech15}: for triphones $A$ and $X$ of the same type, and a triphone $B$ that differs from them only in its centre phone, an error is counted if $X$ is closer to $B$ than to $A$ under DTW with cosine distance.
We report the error rate in the within-speaker (WS) condition, where all three segments come from one speaker, and the across-speaker (AS) condition, where $X$ comes from a different speaker, using \texttt{fastabx}~\cite{poli+etal_arxiv25}.

For the word level, we use the same-different task~\cite{carlin+etal_interspeech11}: all pairs of word tokens of at least 0.5\,s and five characters are compared using DTW, and average precision (AP) measures how well same-word pairs are ranked above different-word pairs.
To isolate speaker effects, we also report same-word same-speaker (SWSP) and same-word different-speaker (SWDP) AP where precision is computed over all word pairs but recall is computed only over same-word pairs from the same or from different speakers, respectively.

\subsection{Features and training}
We use layer 21 of WavLM Large~\cite{chen+etal_jstsp22}, which gave the highest AP among all layers of WavLM and HuBERT on the English development set, in line with the layer-wise analysis of~\cite{pasad+etal_icassp23}.
Features are normalised to zero mean and unit variance per dimension using statistics from each language's training set.

The cAE consists of nine hidden layers of 2048 units each, except for the eighth layer (the bottleneck), which has 128 units.
A linear output layer matches the dimensionality of the input features.
All hidden layers use tanh activations.\footnote{ReLU activations were also tested and performed comparably.}
The encoder $f$ consists of the first eight layers, up to and including the bottleneck, and the decoder $g$ of the remaining layers; $\mathbf{z}$ is taken before the bottleneck activation.
The NCE model uses the same encoder structure with $\tau = \text{0.1}$.

The cAE and NCE model have 29.9M and 27.5M trainable parameters, respectively, compared to 315.5M for WavLM Large, through which no gradients are propagated.
Both models are randomly initialised and trained with Adam~\cite{kingma+ba_iclr15} for 30 epochs with a batch size of 2048 and a learning rate of $\text{10}^{-\text{5}}$.

\subsection{SpidR-Adapt}
We compare against the Reptile variants of SpidR-Adapt~\cite{luthra+etal_acl26}, a speech representation model designed for few-shot adaptation to new domains from small amounts of untranscribed audio.
In contrast to our approach, which keeps the SSL model frozen and trains a separate network on its features, SpidR-Adapt fine-tunes the model, starting from a meta-learned initialisation.
We use two released checkpoints: MAdaPT-Reptile and MAdaPT-Reptile~(SSL/SL).
The latter interleaves phoneme-supervised steps on labelled source-language data.
In both cases, adaptation to the target domain itself is unsupervised.
Both models are Base-sized with 12 layers, 768-dimensional outputs, and 90.1M parameters.
Each model is adapted on the same audio used to train our models, using the released recipe (8,000 steps, learning rate $\text{5} \cdot \text{10}^{-\text{6}}$), except for the batch length which is reduced from approximately 4\,min to 40\,s of audio to fit GPU~memory.

Since SpidR-Adapt has no bottleneck layer, we evaluate all of its layers on the test set of each language and report the one giving the
best overall trade-off between ABX and AP.
This per-language selection on test data gives SpidR-Adapt an advantage over our models, for which selection is fixed.

\begin{table*}[!t]
    \mytable
    \footnotesize
    \setlength{\tabcolsep}{3pt}
    \caption{Within-speaker (WS) and across-speaker (AS) triphone ABX error rates (\%, lower is better) and same-different AP (\%, higher is better) on the test sets, for WavLM features adapted with each method. SpidR-Adapt uses layer 5 for Xitsonga, and layer 6 for English and French. Supervised SpidR-Adapt uses layer 7 for English and Xitsonga, and layer 8 for French.}
    \begin{tabularx}{\linewidth}{@{}l@{\ \ }*{5}{C}@{\quad}*{5}{C}@{\quad}*{5}{C}@{}}
        \toprule
        & \multicolumn{5}{c}{English} & \multicolumn{5}{c}{Xitsonga} & \multicolumn{5}{c}{French} \\
        \cmidrule(lr){2-6} \cmidrule(lr){7-11} \cmidrule(l){12-16}
        & \multicolumn{2}{c}{ABX (\%) $\downarrow$} & \multicolumn{3}{c}{AP (\%) $\uparrow$}
        & \multicolumn{2}{c}{ABX (\%) $\downarrow$} & \multicolumn{3}{c}{AP (\%) $\uparrow$}
        & \multicolumn{2}{c}{ABX (\%) $\downarrow$} & \multicolumn{3}{c}{AP (\%) $\uparrow$} \\
        \cmidrule(r){2-3} \cmidrule(lr){4-6} \cmidrule(lr){7-8} \cmidrule(lr){9-11} \cmidrule(lr){12-13} \cmidrule(l){14-16}
        Model & WS & AS & All & SWSP & SWDP & WS & AS & All & SWSP & SWDP & WS & AS & All & SWSP & SWDP \\
        \midrule
        \multicolumn{16}{@{}l}{\textit{Unsupervised}} \\
        WavLM baseline & 5.35 & 6.91 & 86.0 & 93.9 & 83.2 & 4.21 & 4.88 & 64.9 & 81.5 & 62.8 & 6.41 & 9.82 & 65.8 & 82.2 & 55.2 \\

        cAE (DUSTED pairs) & \textbf{4.32} & \textbf{5.87} & \textbf{93.1} & 95.3 & \textbf{92.3} & \textbf{2.63} & \textbf{3.04} & 82.9 & 86.5 & 82.4 & 5.42 & 7.90 & 72.3 & 82.4 & 65.8 \\
        NCE (DUSTED pairs) & 4.98 & 6.56 & 92.6 & \textbf{95.6} & 91.6 & 3.63 & 4.18 & \textbf{85.2} & \textbf{89.2} & \textbf{84.7} & 5.33 & 7.91 & \textbf{77.3} & 87.4 & \textbf{70.7} \\
        SpidR-Adapt & 5.69 & 7.78 & 83.2 & 93.3 & 79.6 & 2.81 & 3.52 & 80.5 & 88.7 & 79.4 & \textbf{4.18} & \textbf{6.43} & 76.0 & \textbf{91.7} & 65.8 \\
        \midrule
        \multicolumn{16}{@{}l}{\textit{Supervised}} \\
        cAE (gold pairs) & 5.60 & 7.49 & 97.8 & 98.1 & 97.7 & \textbf{2.17} & \textbf{2.58} & 92.0 & 94.5 & 91.7 & 5.61 & 7.95 & 81.6 & 88.6 & 77.0 \\
        NCE (gold pairs) & 6.20 & 7.92 & \textbf{98.2} & \textbf{98.8} & \textbf{98.0} & 3.83 & 4.06 & \textbf{93.2} & \textbf{95.5} & \textbf{93.0} & 5.49 & 8.21 & 83.1 & 91.3 & 77.8 \\
        SpidR-Adapt (supervised) & \textbf{4.93} & \textbf{6.63} & 87.8 & 92.9 & 86.0 & 2.60 & 3.30 & 85.2 & 88.3 & 84.8 & \textbf{3.75} & \textbf{5.51} & \textbf{85.4} & \textbf{92.9} & \textbf{80.6} \\
        \bottomrule
    \end{tabularx}
    \label{tab:main_results}
\end{table*}

\section{Results}

\subsection{Unsupervised adaptation}
The top section of Table~\ref{tab:main_results} shows that both the cAE and the NCE model, trained on DUSTED pairs, improve on the WavLM baseline for every metric and language, without using any transcriptions.\footnote{Repeating experiments across 10 different seeds on English, the standard deviations were at most 0.05\% for ABX and 0.1\% for AP.}
The AP gains stem predominantly from different-speaker comparisons: SWDP AP improves by 8.4--21.9\% absolute,
compared to 0.2--7.7\% for SWSP AP.
This is consistent with the training objectives, which encourage the models to retain linguistic content while discarding speaker-specific information.
The size of the improvement depends on the language.
The largest gains are on Xitsonga, where the cAE improves AP by 18.0\% absolute, compared to 7.1\% on English and 6.5\% on French. It also reduces within-speaker ABX error by 1.6\% absolute on Xitsonga, compared to 1.0\% on both English and French.

French scores are worse than the other languages. This is not
caused by pair discovery alone: when training on gold word pairs from forced alignments (bottom section of Table~\ref{tab:main_results}), AP on French reaches only 81.6--83.1\%, compared to 92.0--93.2\% on Xitsonga, despite the two
having similar baselines.
French may instead be limited by greater speaker and channel variation across its volunteer-read recordings, which is consistent with its large baseline gap between SWSP and SWDP AP.

\subsection{cAE vs NCE}

Comparing our two networks, overall, word-level AP does not consistently favour either model: the NCE model scores better on Xitsonga and French with DUSTED pairs (85.2\% vs 82.9\%, and 77.3\% vs 72.3\%), while the two are comparable on English.
In contrast, the phone-level ABX depends strongly on the training objective: with DUSTED pairs, the cAE reduces ABX error by approximately three times as much as the NCE model on English and Xitsonga, and also has lower ABX error with gold pairs on these languages.
On French, the two objectives perform similarly on ABX.

One theory is that the cAE's reconstruction loss forces it to retain more phonetic detail, which helps
phone ABX.
NCE separates frames from different discovered terms; since negatives could contain the same phone, it may discard some useful phonetic information while still retaining word-level~information.

\subsection{Do we do better with perfect pairs?}

Since our models learn only from discovered pairs, a natural question is whether better pair discovery would give better features.
To test this, we additionally train on gold pairs from forced alignments (bottom section of Table~\ref{tab:main_results}).
Gold pairs improve AP for both models on all three languages, but do not consistently improve ABX; on English, they even make ABX worse than the baseline features.
Despite improvements in word-level discrimination with perfect pairs, the results on phone discrimination therefore make it less clear whether improvements in the term discovery system will necessarily give downstream improvements.

\subsection{SpidR-Adapt}
The comparison with SpidR-Adapt depends on the language.
On English, both of our unsupervised models outperform unsupervised SpidR-Adapt on every metric (e.g.\ AP of 93.1\% and 92.6\%, vs 83.2\%).
On Xitsonga, the cAE obtains the best ABX error (3.04\% vs 3.52\% across speakers) and the NCE model the best AP (85.2\% vs 80.5\%).
On French, however, SpidR-Adapt obtains
better ABX error (6.43\% vs 7.90\% and 7.91\%), with AP worse than the NCE model but better than the cAE model.
The methods differ in their speaker robustness: unsupervised SpidR-Adapt retains a larger gap between SWSP and SWDP AP on every language (9.3--25.9\%
vs 3.0--16.7\%), which is consistent with our pair-based training directly relating different instances of discovered terms.

With supervised pretraining, SpidR-Adapt improves further.
On French it exceeds even our gold-pair models (AP of 85.4\%, vs 81.6\% and 83.1\%), whereas on English its AP of 87.8\% remains worse than that of our unsupervised models.
On Xitsonga, however, our gold-pair cAE model obtains better ABX and AP scores than the supervised SpidR-Adapt.
Notably, our unsupervised cAE performs comparably to supervised SpidR-Adapt on Xitsonga (within-speaker ABX of 2.63\% vs 2.60\%), and has better across-speaker ABX (3.04\% vs 3.30\%), without using any labelled data.

\subsection{Amount of data}
\definecolor{nceOrange}{HTML}{DF6D41}
\definecolor{spidrGreen}{HTML}{AAA648}

\def\spidrLayer{6}

\pgfplotsset{
  dataamount/.style={
    scale only axis,
    width=3.2cm, height=2.9cm,
    xmode=log,
    xmin=8.5, xmax=420,
    xtick={10,30,60,120,234,360},
    xticklabels={10m,30m,1h,2h,4h,6h},
    minor xtick={},
    log ticks with fixed point,
    xlabel={Target-domain speech},
    axis lines*=left,
    tick pos=left,
    tick align=outside,
    ymajorgrids=true,
    grid style={line width=0.3pt, draw=gray!30},
    tick label style={font=\scriptsize},
    label style={font=\footnotesize},
    ylabel style={yshift=-2pt},
    xlabel style={yshift=2pt},
  },
  dataline/.style={
    color=#1,
    line width=1.1pt,
    mark options={solid, fill=#1, draw=white, line width=0.4pt},
    mark size=2pt,
  },
  dataref/.style={line width=0.9pt},
}

\begin{figure}[!t]
  \centering
  \begin{tikzpicture}
    \begin{groupplot}[
      group style={group size=2 by 1, horizontal sep=1.2cm},
      dataamount,
    ]
      \nextgroupplot[
        ylabel={ABX error (\%), lower is better},
        ymin=4.9, ymax=9.3,
        ytick={5,6,7,8,9},
        legend columns=3,
        legend cell align=left,
        legend style={
          at={(1.13,1.06)}, anchor=south,
          draw=none, fill=none,
          font=\scriptsize,
          column sep=3pt,
        },
      ]
      \addplot[baseGrey, dataref, dashed] coordinates {(8.5,6.13) (420,6.13)};
      \addlegendentry{WavLM L21}
      \addplot[dataline=caeBlue, mark=*] coordinates {(10,9.085) (30,7.955) (60,6.230) (120,5.360) (234,5.180) (360,5.095)};
      \addlegendentry{cAE}
      \addplot[dataline=spidrGreen, mark=triangle*] coordinates {(10,6.770) (30,6.755) (60,6.765) (120,6.790) (234,6.750) (360,6.735)};
      \addlegendentry{SpidR-Adapt, L6}
      \addplot[spidrGreen, dataref, densely dotted] coordinates {(8.5,6.595) (420,6.595)};
      \addlegendentry{SpidR-Adapt L6, unadapted}
      \addplot[dataline=nceOrange, mark=square*] coordinates {(10,7.165) (30,6.795) (60,6.275) (120,5.875) (234,5.690) (360,5.770)};
      \addlegendentry{NCE}

      \nextgroupplot[
        ylabel={AP (\%), higher is better},
        ymin=79, ymax=94,
        ytick={80,85,90},
      ]
      \addplot[baseGrey, dataref, dashed] coordinates {(8.5,86.0) (420,86.0)};
      \addplot[spidrGreen, dataref, densely dotted] coordinates {(8.5,82.1) (420,82.1)};
      \addplot[dataline=spidrGreen, mark=triangle*] coordinates {(10,82.6) (30,82.5) (60,82.3) (120,82.3) (234,82.8) (360,83.2)};
      \addplot[dataline=caeBlue, mark=*] coordinates {(10,79.8) (30,83.2) (60,90.2) (120,92.6) (234,92.9) (360,93.1)};
      \addplot[dataline=nceOrange, mark=square*] coordinates {(10,87.9) (30,89.1) (60,90.2) (120,92.1) (234,92.7) (360,92.6)};
    \end{groupplot}
  \end{tikzpicture}
  \caption{ABX error (mean of within- and across-speaker) and AP on the English test set for different
  amounts of target~speech.}
  \label{fig:data_amount}
\end{figure}
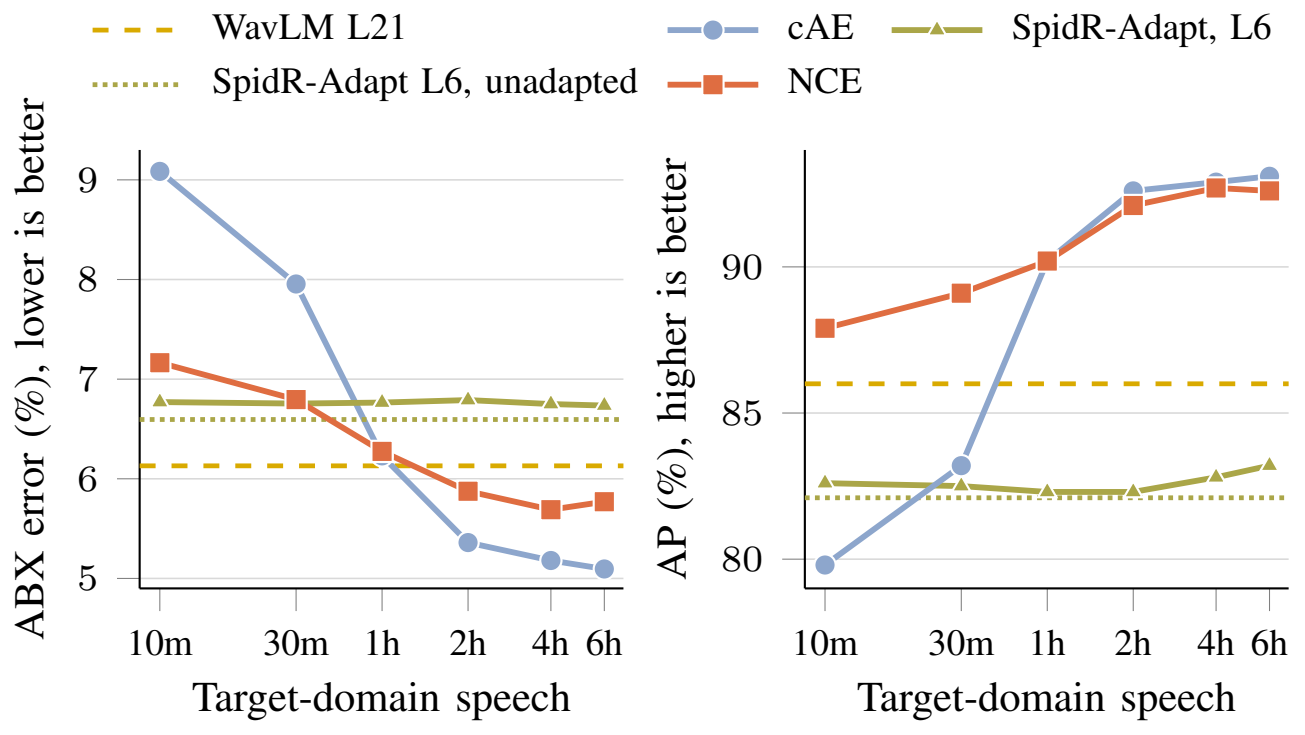

Figure~\ref{fig:data_amount} shows the unsupervised model results on English when each method is given only part of the training speech.
Data is subsampled across the 12 speakers.
Reducing the data from 6 to 2\,h has little effect on the cAE, although the number of discovered pairs falls by 85\%.
Below 1\,h, the cAE is worse than the input WavLM features, since 30 epochs then yield very few updates (90 at 10\,min, vs 10\,560 at 6\,h).
Training for as many updates as the 6\,h run largely recovers it at 30\,min (AP of 90\%), but degrades performance at 1 and 2\,h, so the appropriate training length depends on the amount of data.
The NCE model is more robust, staying above the baseline on AP at every amount.
SpidR-Adapt changes little across this range and stays close to its unadapted scores.

\section{Conclusion}
In this paper we showed that lightweight networks trained on frozen WavLM features can adapt them to a new domain using
spoken-term pairs discovered in an unsupervised fashion.
Both the correspondence autoencoder (cAE) and noise contrastive estimation (NCE) model improved phone-level and word-level discrimination over the baseline WavLM features on English, Xitsonga, and French, with word discrimination
gains coming predominantly from different-speaker comparisons.
The two models have
different strengths: the cAE gave better ABX error on English and Xitsonga, while NCE
gave better word-level average precision on Xitsonga and French.
Compared to SpidR-Adapt, which fine-tunes the model itself, our approach performed better on English and competitively on Xitsonga, but had worse ABX error on French.
When training data was reduced, our methods outperformed SpidR-Adapt and the baseline features when more than 1 hour of data was used.

Our SSL-based pair discovery system currently relies on an English HuBERT checkpoint.
Future work will investigate whether a foundation model matching the target language improves results.
\clearpage
\balance{}
\bibliographystyle{IEEEtran}
\bibliography{IEEEabrv,refs}

\end{document}